\documentclass[sigconf,screen]{acmart}

\AtBeginDocument{%
  }

\title{Developers’ Perspectives on Software Licensing: Current Practices, Challenges, and Tools}

\author[N. Wintersgill, T. Stalnaker, D. Otten, L. A. Heymann, O. Chaparro, M. Di Penta, D. M. German, D. Poshyvanyk]
{Nathan Wintersgill*, 
Trevor Stalnaker*, 
Daniel Otten*, 
Laura A. Heymann*, 
Oscar Chaparro*,
Massimiliano Di Penta\textsuperscript{\dag},
Daniel M. German\textsuperscript{\ddag},
Denys Poshyvanyk*} 
\affiliation{
    \institution{*William \& Mary (USA), \textsuperscript{\dag}University of Sannio (Italy), \textsuperscript{\ddag}University of Victoria (Canada)}
    \country{njwintersgill@wm.edu, twstalnaker@wm.edu, dsotten@wm.edu, laheym@wm.edu, oscarch@wm.edu, dipenta@unisannio.it, dmg@uvic.ca, dposhyvanyk@wm.edu}
}

\date{September 2025}

\usepackage{ifthen}

\newboolean{showcomments}
\setboolean{showcomments}{false}

\newboolean{showboxes}
\setboolean{showboxes}{true}

\newboolean{showrevisions}
\setboolean{showrevisions}{false}

\usepackage{verbatim}

\usepackage[utf8]{inputenc}
\usepackage{url}
\usepackage{colortbl}
\usepackage{balance}
\usepackage{xspace}
\usepackage{graphicx}
\usepackage{bold-extra}
\usepackage{paralist}
\usepackage{multirow}
\usepackage{listings}
\usepackage{boxedminipage}
\usepackage{soul}
\usepackage{enumitem}
\usepackage[normalem]{ulem}
\setlist{nolistsep,leftmargin=.5cm}
\usepackage{booktabs}
\usepackage{tabularx}
\usepackage{svg}
\usepackage{calc}
\usepackage{float}
\usepackage{multirow}
\usepackage[normalem]{ulem}
\useunder{\uline}{\ul}{}
\usepackage[most]{tcolorbox}
\usepackage{caption}
\usepackage{microtype} 
\usepackage{algorithmic}
\usepackage{textcomp}
\usepackage[dvipsnames]{xcolor}

\usepackage{wrapfig}
\usepackage{subcaption}
\usepackage{cleveref} 
\usepackage{adjustbox}
\usepackage{totcount}

\usepackage{breakurl}

\usepackage{hyperref}

\ifthenelse{\boolean{showcomments}}
{\newcommand{\nb}[2]{
		\fbox{\bfseries\sffamily\scriptsize#1}
		{\sf\small$\blacktriangleright$\textit{#2}$\blacktriangleleft$}
	}
	
}
{\newcommand{\nb}[2]{}
	
}

\ifthenelse{\boolean{showrevisions}}
{
\newcommand\rev[1]{{\color{blue} {#1}}}
}
{
\newcommand\rev[1]{{#1}}
}

\newcommand{\ie}{\textit{i.e.},\xspace}
\newcommand{\eg}{\textit{e.g.},\xspace}
\newcommand{\etc}{\textit{etc.}\xspace}
\newcommand{\etal}{\textit{et al.}\xspace}

\newcommand{\ellipsis}{\textit{…}\xspace}

\newcommand{\resp}[1]{$R_{#1}$}
\newcommand{\iresp}[1]{$RI_{#1}$}

\newcommand{\interviewcount}{7\xspace}
\newcommand{\completeresponses}{58\xspace}

\newcounter{findingcounter}
\newtotcounter{totalfindings}
\ifthenelse{\boolean{showboxes}}
{
    \newcommand{\finding}[1]{%
      \refstepcounter{findingcounter}
      \begin{tcolorbox}[boxsep=1pt,left=2pt,right=2pt,top=1pt,bottom=1pt]%
      \small
      \textbf{Finding \arabic{findingcounter}:} #1
      \end{tcolorbox}%
      \addtocounter{totalfindings}{1}
    }
    
  }
{
    \newcommand{\finding}[1]{}
}

\begin{abstract}

Most modern software products incorporate open-source components, requiring development teams to maintain compliance with each component's licenses. Noncompliance can have significant financial, legal, and reputational repercussions. Although some organizations may seek advice from legal practitioners to assist with licensing tasks, developers still play a key role in this process. To this end, it is essential to understand how developers approach licensing compliance tasks, the challenges they encounter, and the tools they use.  
This work studies these aspects of software licensing practices through a study---conducted by a joint team of software engineering and legal researchers---consisting of a survey with \completeresponses software developers and \interviewcount follow-up interviews.  The study resulted in \total{totalfindings} 
key findings regarding the current state of practice. 
We discuss the implications of our findings and offer directions for future research, as well as actionable recommendations.
\end{abstract}

\begin{document}

\maketitle

\section{Introduction}
\label{sec:intro}

Software reuse, particularly through open source software (OSS), has become fundamental in the process of software development, allowing software developers to avoid duplicating efforts in areas where previous developers have found suitable solutions. However, the fact that a component is open source does not mean that it can be used without restrictions. Because software is protected by copyright and patent law~\cite{copyright_software}, 
OSS components are typically released under one or more OSS licenses that govern how a component may be reused, modified, and distributed. Failure to comply with the terms of any applicable license risks liability for infringement, which can result in legal, financial, and reputational consequences~\cite{vizio_1, ai_copyright, johndeere, gunningham2004social, cost1, cost2}, both for individual developers and for the organizations for which they work. 
\looseness=-1

Ensuring license compliance throughout the development process is therefore a critical task. 
However, this process can be challenging to manage. Software systems can involve hundreds or thousands of OSS components, each under a separate license that must be identified and understood~\cite{8728094}. Furthermore, some OSS licenses have conflicting terms, making the use of components under different licenses problematic. There are hundreds of potentially applicable OSS licenses~\cite{spdx_licenses,osi_approved}, some of which use legal terminology that is unfamiliar to developers and that does not always have a broadly accepted interpretation~\cite{almeida2019investigating,gangadharan2012managing,vendome2018distribute}.
\looseness=-1

As a result, it may seem appropriate to offload the license compliance work to legal teams or Open Source Program Offices (OSPOs), but this can delay compliance efforts until near the end of a development cycle \cite{wintersgill2024law}, leading to lower productivity and wasted effort. Thus, while legal experts play an important role in compliance processes, it is also important for developers to be trained in such tasks.
\looseness=-1

\rev{As the individuals closest to the code, developers bear primary responsibility for maintaining compliance with the licenses governing that code~\cite{wintersgill2024law}. Although previous work has identified that licensing tasks can be challenging for developers~\cite{almeida2019investigating,almeida2017software,vendome2018distribute,moraes2021one,german2009license,gangadharan2012managing}, further investigation of developers' perceptions of different licensing tasks is needed, and questions remain regarding developers' use of licensing tools and the strategies employed to address licensing tasks.}
\looseness=-1

\rev{To fill this knowledge gap, this study surveyed software developers with various levels of familiarity with licensing tasks. Given that licensing tasks might not need to be performed as frequently as other software engineering tasks and that developers do not always have training regarding software licensing, this study provides important context by representing the views of developers with realistic licensing workloads and training, spanning those with limited exposure to licensing tasks to those who address such tasks frequently, which results in a novel view of the state of practice.} %
Our goal was to gain a deeper understanding of licensing processes and challenges for developers, discover the gaps in the technology developers use for licensing, and work toward organizational and technical solutions. We accomplished this by identifying the licensing tasks that developers perform most frequently and the tasks that are most difficult, explaining how developers address licensing tasks through strategies that encompass both communication with legal experts and making changes themselves, and diagnosing low awareness and adoption of tools that can assist with licensing tasks.

\rev{

The main contributions of this paper are the following:
\begin{itemize}
    \item{A novel assessment of how developers not only understand but also engage with software licensing tasks, based on a survey of \completeresponses developers with various levels of experience in licensing;}
    \item{An in-depth analysis of the processes, roles, and tools that contribute to addressing software licensing issues, including the adoption of available tools and availability of resources in different contexts; and}
    \item{Actionable insights into where future licensing tooling development should be directed, including key pain points and opportunities for generative AI.}
\end{itemize}

}

\section{Background and Related Work}
\label{sec:back}

\subsection{Software licensing}
As software is protected by copyright law (among other legal doctrines) in the United States and around the world~\cite{copyright_software}, the owner of a piece of software's copyright holds the rights to distribute, reproduce, and create derivative works of that software, as well as the right to authorize others to do so~\cite{UScode17sect107}. Such authorization often occurs by means of a license, which can grant permissions such as using, modifying, or redistributing the software, as well as potentially applying restrictions to those permissions.~\cite{meeker2017open} Use outside of the scope of the license constitutes copyright infringement, unless such use constitutes a fair use under U.S. copyright law.

OSS developers often choose one of several existing popular licenses to apply to their software, such as the MIT~\cite{mit} license or a member of the GPL~\cite{gpl_license} family of licenses. Permissive licenses, as the name suggests, typically allow broad use of the software, requiring only that, for example, the user provide notice files with information on the license or on attribution. Copyleft licenses, by contrast, constrain future use of the software by, typically, requiring that the user make the full source code of their work available under the same license. A developer must therefore understand, at a minimum, which licenses apply to components they incorporate into their work and what those licenses require if they wish to remain compliant and avoid consequences of noncompliance.

Although several licenses have been widely used for many years, interpretive questions still exist. Besides the preliminary questions about the types of use a license permits or prohibits, questions can also arise about whether two licenses are compatible with each other, 
as well as how certain terms apply to desired software uses. For example,  the requirements of GPL v2 apply to any work ``based on the Program," which the license defines as equivalent to ``any derivative work under copyright law." But this language does not answer the question of whether, for example, system calls to a library create a work ``based on the Program" or whether the answer depends on whether the link is static or dynamic, a question that even OSS lawyers have identified as unresolved ~\cite{wintersgill2024law}.

\subsection{Developers performing licensing tasks}

Previous work has shown that software developers often find licensing tasks difficult, such as determining whether two licenses are compatible with each other~\cite{almeida2019investigating,almeida2017software}, selecting or changing licenses for their software~\cite{vendome2015and}, ensuring that their software complies with all relevant licensing requirements~\cite{vendome2018distribute}, and releasing their software under multiple licenses~\cite{moraes2021one}. \rev{Building on this work, we seek to contextualize these challenges by examining how often they are encountered and how developers address them.}

\rev{Both open source and proprietary tools can help with license compliance, such as Fossology~\cite{Fossology} and Black Duck~\cite{BlackDuck}, respectively. Software bills of materials (SBOMs) can provide insights into software's composition, facilitating licensing tasks~\cite{biSbom,stalnaker2024boms,xia2023empirical}. Additionally, }researchers have developed a variety of tools and processes aimed at helping developers with compliance tasks~\cite{tuunanen2009automated, ombredanne2020free, kapitsaki2017automating, german2009license, german2012method, german2010understanding,hemel2011finding, feng2019open,gangadharan2008license}. Some of these studies have explored the use of generative AI to assist in compliance tasks~\cite{cui2025exploring, kahol2025oss, ke2025clausebench}, although the effectiveness of such tools still needs to be further evaluated.
Other studies, based on mined software repositories and Q\&A websites, have explored methods to detect non-approved licenses, non-standard license variants, and exceptions~\cite{meloca2018understanding, zacchiroli2022large, papoutsoglou2022analysis, vendome2017machine, xu2023understanding}.

Prior work has explored some of the decisions developers make during license compliance efforts.  \textit{Wu} \etal conducted a large empirical study aimed at understanding license usage~\cite{wu2024large}. Works by \textit{Vendome} \etal and \textit{Di Penta} \etal sought to understand when, how, and why developers opt to change the license of their software~\cite{di2010exploratory, vendome2017license}.  \textit{Liu} \etal introduced a method to predict license changes from source code changes~\cite{liu2019predicting}.  Our previous work surveyed legal professionals with experience in software license compliance to understand their challenges and their views on software license compliance tasks~\cite{wintersgill2024law}.  Lastly, \textit{Li} \etal provide a systematic literature review of stakeholders' challenges in license compliance tasks~\cite{li2025open}. 
\looseness=-1

\rev{Whereas the literature offers insights, solutions, and tools for licensing tasks, this work seeks to expand on previous efforts in two key ways: (1) by understanding if and how developers adopt these tools and techniques in practice, and (2) by identifying the license compliance pain points that are most relevant to developers by providing greater insight into the specific problems developers report encountering most often, the problems they report as being most difficult, and the strategies they use to address them.}

\vspace{-10pt}

\section{Study Methodology}
\label{sec:methodology}

The \emph{goal} of this study is to analyze the processes employed and challenges faced by software developers in performing OSS license compliance tasks %
to facilitate an understanding of the areas developers struggle with most and enable targeted work developing solutions to such pain points. 
The study was conducted through a survey and follow-up interviews and involved a collaboration between software engineering (SE) 
and law researchers. The \emph{context} of the study consists of \completeresponses developers recruited through personal contacts and open-source projects' discussion channels.

The study aims to address the following research questions (RQs):

\begin{enumerate}[label=\textbf{\labelitemi \space RQ$_\arabic*$:}, ref=\textbf{RQ$_\arabic*$}, wide, labelindent=5pt, leftmargin=5pt]\setlength{\itemsep}{0.2em}

    \item \label{rq:1}{\textit{Which licensing tasks do developers struggle with?}}  This RQ explores the licensing tasks developers encounter most frequently and which are the most difficult to complete.
    
    \item \label{rq:2}{
    \textit{What are the processes by which software licensing problems are resolved?}} 
    This RQ explores how developers 
    resolve compliance issues when they occur and investigates
    how developers perceive the relationship between development and legal teams.
    
    \item \label{rq:3}{\textit{How are developers incorporating license compliance tools in their workflows, if at all, and what issues have they encountered?}} This RQ explores how licensing tools are used by developers, if at all, and issues developers report when using such tools.

\end{enumerate}

\begin{figure}[t]
    \centering
    \includegraphics[width=\linewidth]{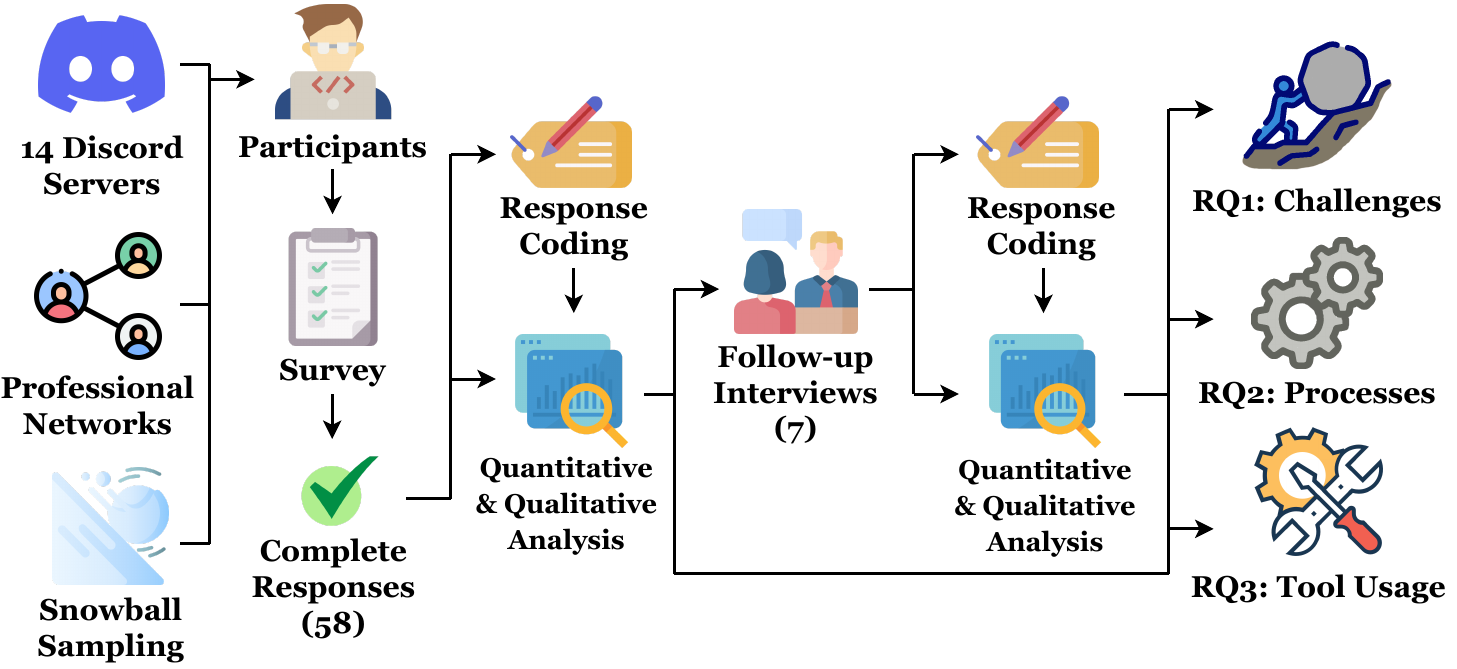}
    \caption{Research methodology (image credits at \cite{anonymous_repo})} %
    \label{fig:methods}
\end{figure}

This study, including the survey questionnaire, the participant identification procedure, and the survey and interview protocols, was approved by our institution's ethics review board. 
An overview of our methodology is depicted in \Cref{fig:methods}.

\subsection{Survey design and participant identification}

\begin{figure*}[t]
    \centering
    \includegraphics[width=0.9\linewidth]{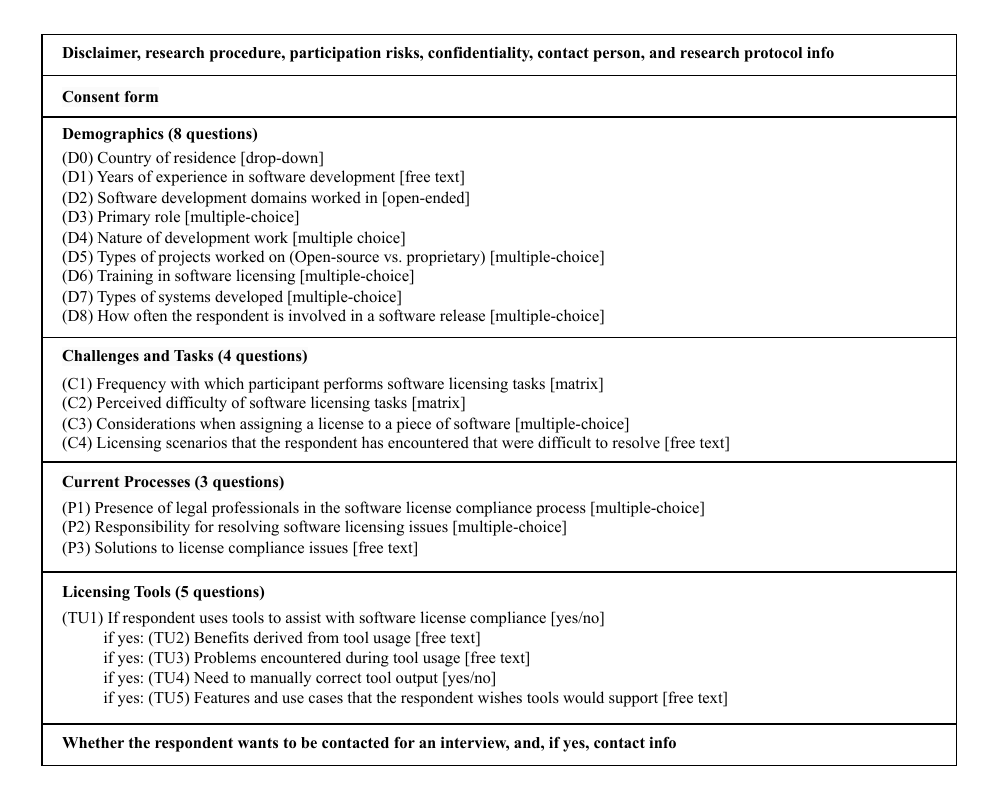}
    \caption{Survey questionnaire outline} %
    \label{fig:outline}
\end{figure*}

When designing our survey questionnaire, we considered previous literature on software license compliance, as well as general guidelines \cite{survey} and SE-specific guidelines for survey design \cite{DBLP:journals/sigsoft/PfleegerK01,DBLP:journals/sigsoft/KitchenhamP02,DBLP:journals/sigsoft/KitchenhamP02a,DBLP:journals/sigsoft/KitchenhamP02b,DBLP:journals/sigsoft/KitchenhamP02c,DBLP:journals/sigsoft/KitchenhamP03}. The survey went through multiple review cycles, involving seven SE researchers and one law researcher, to achieve appropriate content, provide clear and concise language, and eliminate confusion and bias. We tried to limit the response time of the survey to 10 minutes or less to minimize the cognitive load on the participants and to encourage higher completion rates.  In addition, we conducted a small pilot study with members of our research laboratory to further refine the survey questions\rev{, including by identifying questions that could be culled or reworked to better focus on the study's research goals}. The final survey included questions to obtain information related to each of our RQs, including questions relating to the compliance challenges faced by developers, existing compliance practices, and the use of tooling or other automation to perform compliance. %
 \Cref{fig:outline} provides an overview of our survey questionnaire. The complete survey text and information necessary for analysis can be found in our online replication package~\cite{anonymous_repo}. \rev{(Compliance with our ethics board's requirements prevents us from supplying all data collected.)}

\rev{To recruit developers with diverse experiences, we distributed the survey (1) through our professional networks (posts on LinkedIn and emails), (2) through snowball sampling, encouraging respondents to share the survey within their networks, and (3) by posting links to the survey on relevant servers on Discord.}
\rev{We considered various other options for survey distribution (\eg posting on relevant subreddits), but the selected methods allowed us to reach relevant communities while mitigating the risk of receiving responses from people outside the target population, which might occur more from survey invitations posted in highly public forums.} 

To identify Discord communities, we examined the TIOBE index~\cite{tiobeIndex} of top programming languages to identify the top 20 most popular programming languages as of January 2025. We then searched for Discord servers associated with these languages (excluding Scratch, which is primarily a tool for beginners) by locating server links through the languages' own documentation or websites, through third-party Discord server browsing tools DiscordMe~\cite{discordme} and Disboard~\cite{disboard}, and through posts on subreddits related to the targeted languages. This process was supplemented by adding three generalized programming/development-focused servers and one server each for the popular React and Node.js frameworks. This yielded 27 Discord servers representing 16 of the top programming languages as well as general programming spaces and popular frameworks. Next, one researcher carefully analyzed the rules of each Discord server to determine whether it would permit a solicitation for participation in a survey, asking server administrators in cases where an answer was not immediately clear. In total, we were able to post survey invitations in 14 software development-focused Discord servers, with total membership exceeding 485,000 users at the time of distribution. Two additional posts were made in smaller servers but were later removed by moderators. %

\subsection{Survey response collection and analysis}

Survey responses were collected using Qualtrics~\cite{qualtrics} from February 2025 to June 2025. Survey invitations were sent out iteratively during this same period.  %
\rev{To monitor where responses originated from, we distributed survey URLs with identifiers unique to each source of participants.}
We obtained \completeresponses complete survey responses.
\rev{Of these, 38 came from our professional networks and 17 came from Discord, while 3 participants seemingly removed the tracking component from the survey URL. The largest population from a single Discord server was 6 participants.} 

Responses to the five open-ended questions were analyzed through a qualitative coding approach \cite{spencer2009card}.  Three authors, all SE researchers, performed \textit{open-coding} by independently assigning one or more \textit{codes} to each response using a shared codebook (\ie a set of spreadsheets).  Each annotator independently coded all \completeresponses responses, adding new codes to the codebook as necessary.  Once the initial coding was completed, the annotators met to settle disagreements and consolidate the set of codes \rev{by discussing each response for which disagreements occurred and mutually deciding on a final set of codes to apply}. Reasons for disagreement included different interpretations of potentially ambiguous responses, applying the same codes in different contexts, and using broader codes rather than more specific ones. All such disagreements were resolved among the annotators, resulting in a final set of codes, definitions, and assignments for each response. Our replication package~\cite{anonymous_repo} contains our final codes and definitions. We did not base our analysis on inter-rater agreements, as the nature of the study was such that codes were developed through an inductive process, and multiple codes could be assigned to each survey response. However, we carefully followed the best open-coding practices \cite{spencer2009card}, and we leveraged coders' discussions to ensure the reliability of the results.
\looseness=-1
 
We corrected grammar and spelling errors in quotations presented in~\Cref{sec:results} for readability and indicated by ellipses or brackets any deletions or additions to text for clarity or space-related reasons. All responses have been attributed to survey participants using anonymous IDs for traceability (\eg \resp{12}).

\subsection{Interview design and analysis}

To better understand the responses of the participants and gain further insight into their views on software licensing and related issues, we conducted semi-structured follow-up interviews over Zoom. At the end of the survey, we provided respondents with the option to leave an email address if they were willing to participate in a follow-up interview\rev{, which 21 participants utilized. We contacted all 21 of these individuals; of those,  \interviewcount participants responded and agreed to participate in follow-up interviews.}

Each interview lasted 20 to 30 minutes and was recorded to facilitate analysis. The recordings were transcribed using Whisper~\cite{whisperRepo} and reviewed against the original recordings for accuracy by each researcher. Three reviewers assigned topic labels to 
\rev{the interviewee's answers in}
each transcript, and
all three reviewers then discussed the labels to reach a consensus.

We have assigned each interviewee an anonymous identifier to attribute the source of quotations (\eg \iresp{4}). As with the survey results, we have corrected grammar errors for readability and indicated deletions or additions by ellipses or brackets.
\looseness=-1

\subsection{Participant demographics}
\label{sec:demographics}
\rev{Respondents came from 21 countries, with the largest representation from the United States (22, 38\%), Germany (6, 10\%), Bangladesh (5, 9\%), and Colombia (5, 9\%). }
Forty respondents (69\%) reported that they \rev{have} worked on proprietary software, 33 (57\%) on open-source projects that relied on third-party components, and 14 (24\%) on open-source projects with no dependencies. \rev{The presence of these different groups allows us to capture information on licensing tasks for different use cases.} %
Most of the respondents described their primary role as programmer (32, 55\%), with other common roles including researcher (10, 17\%), project technical lead (5, 9\%), tester (3, 5\%), educator (3, 5\%), and software architect (2, 3\%).  Development experience ranged from less than one year to 35 years, with an average of 9.9 years and a median of 7 years. %
Domains also varied broadly, ranging from contracted government work and healthcare to game development and banking. Almost two-thirds of the respondents had experience with web development (37, 64\%). The respondents also developed libraries/frameworks (25, 43\%), desktop applications (20, 34\%), mobile applications (15, 26\%), development tools (14, 24\%), middleware (13, 22\%), and AI-intensive systems (9, 16\%).  Although not extensive, the developers in the response pool represent a range of (self-reported) experiences and backgrounds, likely representing diverse experiences with software licensing.

We evaluated participants' \rev{background} with licensing by asking if they had any training or education in software licensing, and, if so, whether that training was formal (\eg college classes, degree, or certification) or informal (self-learning, employer-provided, \etc). 
\rev{Thirty-five (60\%) of our respondents reported that they had some kind of training. In total, 25  (43\%) reported informal training, 4 (6.9\%) reported formal training, 6 (10.3\%) reported both formal and informal training, and 23 (39.7\%) reported no licensing training.}
\rev{Despite low reported amounts of formal licensing training, almost all of these developers, as we will show in~\Cref{sec:results},  have engaged in licensing tasks in some way.

As our previous work indicates, developers are on the front lines of license compliance~\cite{wintersgill2024law}. This current work captures developers with a variety of backgrounds and experience levels to provide a realistic view at the current state of practice.}

\section{Results}
\label{sec:results}
In the following, we present the results of our study, which emerged from the survey responses and interviews.

\subsection{\ref{rq:1}: Developer Licensing Tasks}

We discuss participants' responses regarding the different software licensing tasks they perform and the challenges they face. 

\subsubsection{Most common tasks}
\label{sec:common}

\begin{figure}[t]
    \centering
    \includegraphics[width=\linewidth]{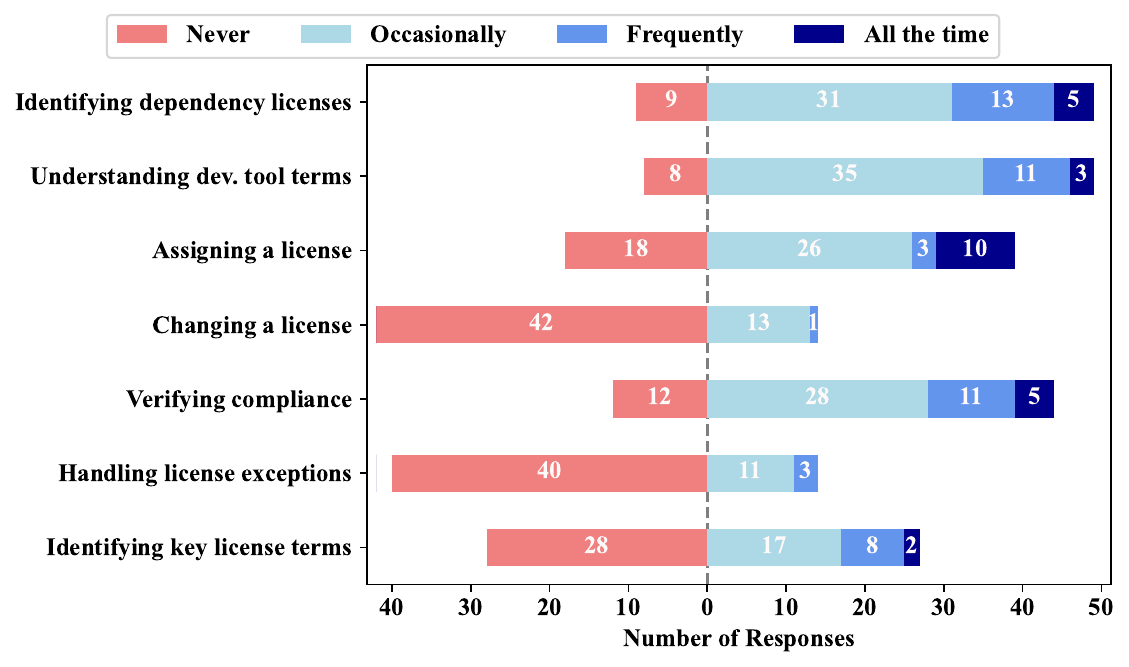}
    \caption{How frequently developers perform licensing tasks}%
    \label{fig:task_frequency}
\end{figure}

Participants were presented with a list of software licensing-related tasks and asked to identify how often they participated in each task on a scale of ``never,'' ``occasionally,'' ``frequently'' or ``all the time.'' \rev{The tasks were derived from our previous experience working with \textit{licensing issues} (such as license incompatibilities, license noncompliance, and problems with distribution) and from our prior work cataloging licensing bugs~\cite{vendome2018distribute}.} 
\rev{Notably, 55 respondents (95\%) reported performing at least one licensing task at least occasionally. This suggests that most developers encounter licensing tasks at some point, further emphasizing the need to understand where the difficulties lie in these tasks. } As shown in \Cref{fig:task_frequency}, the tasks that developers reported performing the most frequently (\ie frequently or all the time) were identifying the licenses of the software components they depend on, understanding the terms of use/service associated with the tools they use to develop software, and verifying that their software complies with all relevant licenses.  
Developers reported that they assigned a license to their own software slightly less often, though this was the task most reported by the respondents as being performed ``all the time'' (10). By contrast, most of the respondents indicated that they never spent time changing their software's license (42) or acquiring or granting license exceptions (40), and slightly less than half reported that they never spent time identifying key license terms (28). %
While almost all developers reported performing licensing tasks at some point, all tasks were reported as being performed ``never'' or ``occasionally'' more than they were reported being performed ``frequently'' or ``all the time.'' This suggests that while most developers will need to address licensing tasks sometimes, developers perform licensing tasks relatively infrequently overall.

\finding{While developers broadly reported that licensing tasks may not be common, the ones they engage with most are identifying their dependencies' licenses and understanding the license terms associated with development tools, as well as verifying that their software is in compliance with all relevant licenses and assigning a license to software.}

Changing a software project's license can be difficult~\cite{wintersgill2024law}, particularly for open-source projects, but such changes occur \cite{vendome2017license}. Few developers, however, reported having experience with changing the license of a software component, and no respondents reported performing this task ``all the time.'' Similarly, most developers reported that they had no experience in acquiring or granting license exceptions, in which an exception to the terms of a component's license is granted to a particular licensee (for example, to grant additional permissions or remove restrictions that would otherwise be in place). License exceptions can be difficult to obtain from open source projects~\cite{wintersgill2024law} and may be a process with which developers are unfamiliar; indeed, this category received the most ``I don't know'' responses, as shown in \Cref{tab:idk}. 

Almost half of the respondents reported that they never spent time identifying key terms from licenses. This may suggest that developers spend little time analyzing the contents of software licenses, or that they feel they understand available software licenses and how they interact, so they do not need to spend further time reviewing them.

\finding{Most developers reported never changing their software's license or obtaining/granting license exceptions. Almost half of developers never dedicate time to identify the key terms of software licenses.} %

\subsubsection{Most difficult tasks}
\label{sec:difficult}

\begin{figure}[t]
    \centering
    \includegraphics[width=\linewidth]{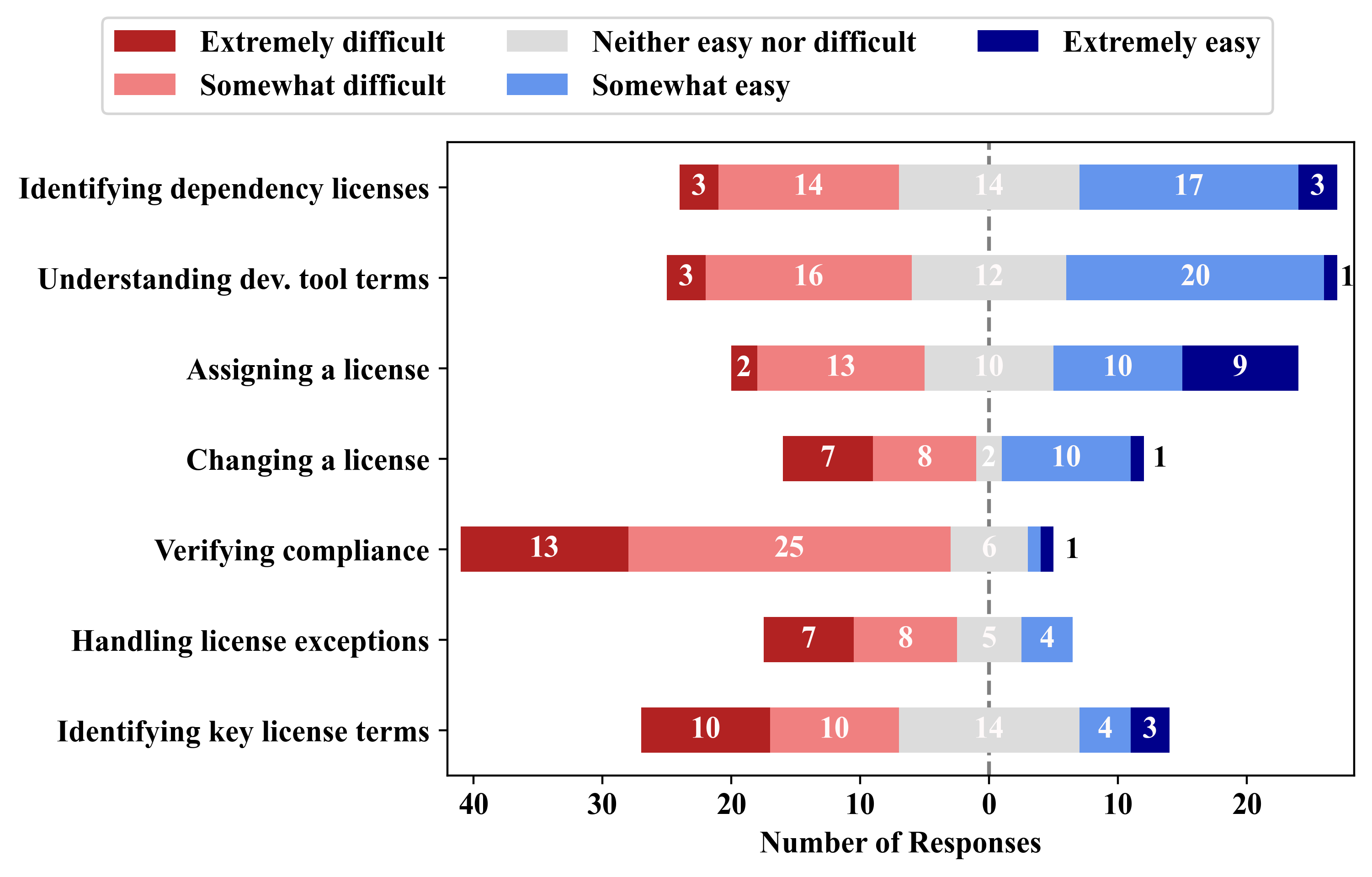}
    \caption{Developers' perceived difficulty of licensing tasks}
    \label{fig:task_difficulty}
\end{figure}

Developers were also asked to rate the difficulty of licensing tasks using a 5-point Likert scale from ``extremely difficult'' to ``extremely easy.'' As shown in \Cref{fig:task_difficulty}, we observed that tasks such as identifying dependency licenses, understanding terms associated with development tools, and assigning a license were viewed as easiest, while changing a license, dealing with license exceptions, and identifying key license terms received more mixed responses. 
\rev{Notably, as \Cref{fig:task_difficulty} shows, participants did not broadly accept any task as being easy: the results show mixed opinions at best for most of the tasks examined. These results confirm previous work showing that developers can find licensing tasks challenging~\cite{almeida2017software, almeida2019investigating}}
\rev{while highlighting that some issues are viewed as being particularly difficult. }

Verifying whether software complies with all relevant licenses was cited as the most difficult task by respondents, being the task most frequently labeled ``somewhat difficult'' (25) as well as most frequently labeled ``extremely difficult'' (13). This is particularly notable, as compliance verification was reported as one of the most frequently performed tasks in the previous question. %
Identifying key license terms was also frequently cited as somewhat/extremely difficult. 
These trends in perceived difficulty are present across both developers who answered that they perform one or more licensing tasks ``frequently'' or ``all the time'' as well as developers who did not. Similarly, those who performed licensing tasks more frequently did not identify any tasks as significantly easier than those who did not perform licensing tasks frequently.

\rev{That developers struggle to understand licenses and their terms, and that verifying compliance proves a difficult task for them, is widely known in this area~\cite{almeida2019investigating, gangadharan2008license, gangadharan2012managing, vendome2018distribute}.} 
\rev{Our interviews provide additional context in this matter. For example, \iresp{1} details how different types of software can be more difficult to manage from a licensing perspective: ``In order of [compliance] difficulty, you'd have the organizationally supported open source is the easiest, then proprietary software, then the non-organizationally supported open source because that one typically lacks consistent documentation, like consistent versioning, [\etc].'' \iresp{1} also contextualizes the difficulty in identifying licensing terms by explaining the practicality of it: ``I find it difficult\ellipsis because of the length of licenses. If they're not using a predefined license, like the Apache licenses or things like that [...] it can be difficult, just because it's  several pages of licensing terms.'' Beyond parsing many terms, it can also be necessary to interpret these terms as they apply to one's specific circumstances. \iresp{7} identifies that ``it's very hard to make sure that your interpretation is the right one. And since we are talking in an international context, sometimes [...] there are some cultural aspects that make you interpret the law in some ways. But when you are talking about something which is worldwide, it's very difficult to interpret how the people that designed the rules thought about it.''}

\finding{Verifying license compliance, while one of the most common licensing tasks, was also reported to be the most difficult. Identifying key license terms was also frequently perceived as difficult.}

We also note that a sizable group of respondents answered ``I don't know/no opinion'' to the survey questions regarding licensing task frequency and difficulty, as shown in \Cref{tab:idk}. In particular, we highlight the large number of ``I don't know'' answers regarding the difficulty of changing a license (30) or receiving/granting license exceptions (34), corresponding with similarly high numbers of respondents who reported that they never performed these tasks and indicating a lack of familiarity with them among developers.

\begin{table}[]
\begin{tabular}{lcc}
\hline
\multicolumn{1}{c}{\textbf{Task}} & \multicolumn{1}{c}{\textbf{How often}} & \multicolumn{1}{c}{\textbf{How difficult}} \\ \hline
Identifying dependency licenses   & 0                                                    & 7                                                                     \\
Understanding dev. tool terms     & 1                                                    & 6                                                                     \\
Assigning a license               & 1                                                    & 14                                                                    \\
Changing a license                & 2                                                    & 30                                                                    \\
Verifying compliance              & 2                                                    & 12                                                                    \\
License exceptions                & 4                                                    & 34                                                                    \\
Identifying key license terms     & 3                                                    & 17                                                                    \\ \hline
\end{tabular}
\caption{``I don't know'' / ``I have no opinion'' responses to survey questions about task frequency and difficulty}
\label{tab:idk}
\end{table}

\subsubsection{Licensing factors}
\label{sec:licensing_factors}

\begin{figure}
    \centering
    \includegraphics[width=\linewidth]{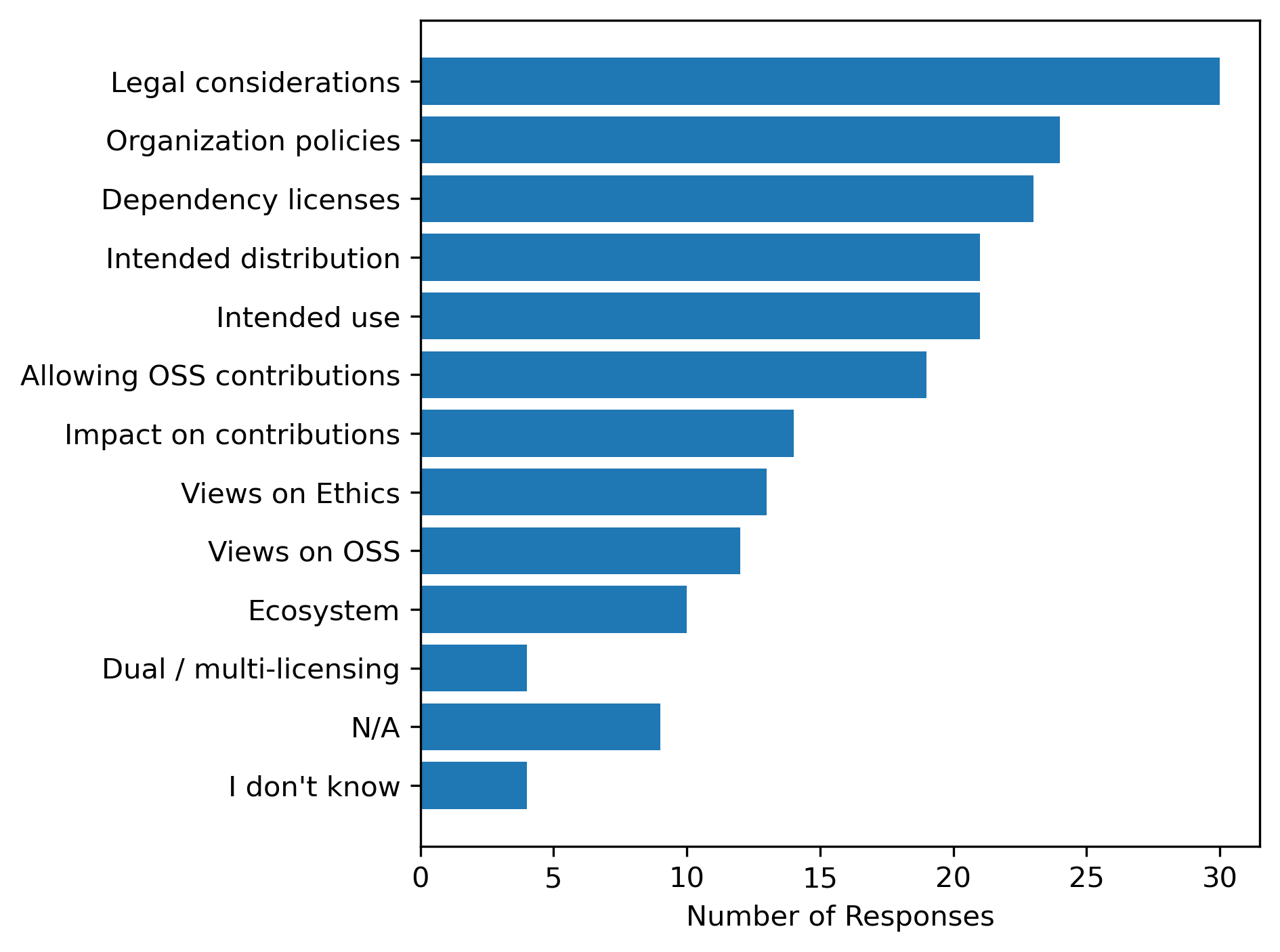}
    \caption{Factors considered when assigning a license}
    \label{fig:licensing_factors}
\end{figure}

To better understand the elements involved in the decision to license software, participants were asked to identify the factors they or their organizations consider when assigning licenses to software. Participants could select multiple such factors. As shown in~\Cref{fig:licensing_factors}, developers most commonly cited legal considerations surrounding their software (30) as influencing the license assigned to their software. Together with the second most popular answer, organizational policies (24), this illustrates how licensing decisions are often made for non-technical reasons. 
The third most frequent answer, dependency licenses (23), shows how constraints from third-party components similarly affect such decisions. Only beyond this point do we observe specific considerations related directly to development-focused concerns, such as the software's intended distribution (21) or intended use~(19). Philosophical considerations, such as the views of the participant or their organization on ethical software distribution and use~(13) or their views on the nature or purpose of OSS development~(12), received fewer responses. 
\looseness=-1

Overall, these results highlight a hierarchy of considerations when choosing a license: first, any legal obligations or organizational policies must be adhered to, then technical limitations based on the project and how it should be used, followed by philosophies regarding software and its reuse. It should be noted that, as participants were able to select multiple responses, a given respondent could highlight multiple of these groupings to be considered in tandem with each other, and for some respondents, the categories may overlap. However, the preeminence of non-engineering or non-usage requirements illustrates the first priorities respondents reported having when making licensing decisions. \rev{As such policies may vary among jurisdictions or among organizations, we suggest that licensing tools should accommodate organizational choices.}

\finding{When considering how to license software, the surveyed developers appeared to first consider legal and organizational obligations before applying technical constraints based on the project and then, less commonly, their philosophy on licensing or OSS. 
}

\subsection{\ref{rq:2}: Licensing Issue Resolution}

We discuss how study participants reported resolving licensing problems, including the involved roles, the adopted resolution process, and how the participants understand licenses.
\looseness=-1

\subsubsection{Roles}
\label{sec:roles}

Legal professionals often participate in licensing compliance tasks, either in house or as outside counsel~\cite{wintersgill2024law}. We asked participants whether they or their organization regularly worked with legal professionals or people otherwise trained in software licensing and/or copyright law. Eighteen participants responded ``Yes,'' 25 responded ``No,'' and 15 responded ``I don't know.'' 
This indicates that expert-level licensing knowledge %
may not always be available to developers (or %
they may not be aware of it). This creates a challenging situation, since developers often do not have training in software licensing and find licensing tasks challenging~\cite{almeida2019investigating}.

\begin{figure}[t]
    \centering
    \includegraphics[width=\linewidth]{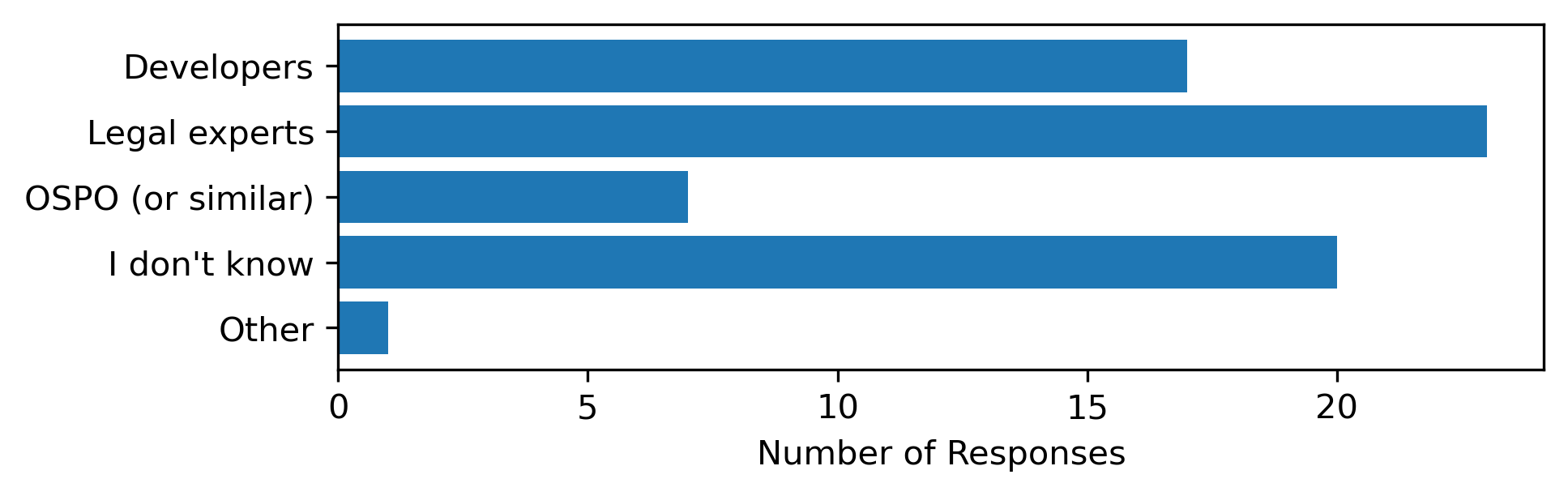}
    \caption{Roles perceived by developers as responsible for addressing licensing issues}
    \label{fig:responsible_for_licensing}
\end{figure}

To understand developers' perspectives on the issue, we asked participants, 
``If an issue related to software licensing arises during software development, who is responsible for resolving that issue?'' Participants could select developers, legal experts, OSPO, or enter another role under ``other.''  
As shown in \Cref{fig:responsible_for_licensing}, the respondents were divided on the issue. Although legal experts were listed most frequently (23), developers were also often chosen (17). 
Notably, 5 respondents who answered that they did not regularly work with legal professionals and 4 respondents who reported not knowing if they or their organization regularly worked with legal professionals nevertheless selected legal experts as responsible for resolving licensing issues, indicating that such answers were likely based on respondents' beliefs regarding these responsibilities more than lived experience or known procedure. The participants' answers may have depended on the resources available to them. Of the 40 respondents who indicated that they have worked on proprietary software, 20 said legal experts were responsible, while of the remaining 18 respondents who did not work on proprietary software, only 3 listed legal experts as responsible. Additionally, many respondents indicated that they did not know whose responsibility it was (20), which may indicate limited experience on this topic. 
Developers rarely cited an OSPO~\cite{munir2021rise}, a group dedicated to managing activities related to OSS,
as responsible for addressing licensing issues (7). 
One respondent selected ``other'' and elaborated that they were solely responsible for licensing in their case. %

The number of respondents who viewed legal experts as responsible for addressing licensing issues (23) exceeds the number of developers who indicated that they regularly worked with such experts (18). 
This may indicate that the processes for addressing licensing issues are not standardized and may not meet stakeholder expectations. We discuss the implications of this in \Cref{sec:who_responsible}.

\finding{Developers are divided on who is responsible for addressing licensing issues. Similarly sized groups chose legal experts and developers as responsible, while just as many did not know who was responsible.}

\subsubsection{Problem resolution process}
\label{sec:process}

To gauge the difficulty that developers had when addressing licensing issues, we asked survey participants to answer whether solutions to licensing issues were typically immediately apparent or if such issues required research. The participants were invited to elaborate. Of the 33 respondents who indicated that they had experienced licensing issues, only 6 respondents (10.34\%) reported that solutions were readily apparent. Twenty-one respondents (36.21\%) indicated that addressing licensing issues required additional research, while 6 respondents (10.34\%) said that it depended on the nature of the problem. 

Many respondents then provided details on how they might address such problems. The most common such recourse was to involve a legal team in the process, as specified by 9 respondents (15.52\%). Sometimes, this was represented as an interactive process involving both lawyers and developers: \resp{33} noted that ``[t]he complexity depends on the type of license, usage rights, and legal implications, so it usually involves reviewing the specific terms of the license and possibly consulting legal or compliance experts.'' In other cases, the process involved less ``consulting'' in favor of passing the issue to others, such as \resp{40}, who said that issues went to ``[m]y manager, prior [to contacting] the legal advisory team,'' and \resp{38}, who noted that ``[f]urther investigation is passed to an expert in corporate law.'' \rev{However, as noted above, legal professionals may not always be available to developers. Given that this was the most-reported strategy, this demonstrates a need for more accessible options for addressing licensing issues, or, at least, better adoption of such strategies.} Four respondents (6.90\%) stressed that an understanding of the involved software licenses and their terms or requirements is a necessary prerequisite to solving licensing problems. As \iresp{2} indicated in their survey response, to find solutions to licensing issues, ``you need to fully understand licensing in the first place.''
\looseness=-1

\finding{When faced with licensing issues, developers' most-reported strategy was escalating the issue to legal experts, but solutions often require additional research from developers and depend on the problem.}

The respondents also reported several specific strategies that could be deployed once a licensing issue was identified. Three respondents (5.17\%) indicated that when issues with a dependency's license arise, they might search for an alternative component with similar functionality. Two of these (3.45\%) further noted that, as a last resort, they might write an internal alternative if a third-party solution was not feasible for licensing reasons. \resp{14} explained: ``Usually we pivot to a different dependency with a compatible license or ask for license exceptions if possible. We had one case where we had to resort to doing an [in-house] development because the only library that matched the requirements was incompatible.''

\iresp{6} noted that, concerning the timing of licensing tasks in the software development cycle, ``the bulk happens at the first release.'' This aligns with prior work showing that many organizations treat licensing as something to be addressed at the end of development, although such an approach introduces new challenges~\cite{wintersgill2024law}.

Solutions that appeared only once (1.72\%) in our results  
included reverting to an older version of a dependency if that version had a more favorable license, purchasing a license when available, simply ceasing use of the affected component, or letting the issue go unaddressed. 
\rev{While the methods identified here align with previous work~\cite{xu2023understanding}, our results show that no method was particularly favored, highlighting the need for licensing solutions and, by extension, tools that can offer solutions tailored to the specific circumstances of the software in question.}
\looseness=-1

\finding{Developers described disparate strategies for addressing licensing issues at the dependency level, including changing dependencies, reverting to an old version, and implementing functionality themselves, demonstrating a lack of uniform practice for handling licensing issues.}

\subsubsection{Understanding licenses}

\rev{In interviews, we sought to clarify how developers acquired an understanding of software licenses.} 
\iresp{3}, who had indicated that they had received informal license training \rev{(see \Cref{sec:demographics})}, described a ``bootcamp'' for licensing consisting of ``three sessions of 45 minutes.'' Some interviewees described how they had gained an understanding of common licenses over the years, such as \iresp{5}, who noted, ``I generally know that if it's MIT or Apache, I'm good to go ... and those are the ones that I encounter most regularly.'' \iresp{6} had a similar familiarity with their most-used licenses but acknowledged that this did not extend to licenses they had less experience with: ``I'm familiar with ... the [Creative Commons] sharealike, the MIT license.... But I wouldn't say that I'm comfortable knowing what all the most common ones mean. I know the ones that I know because they are the ones that I use.'' \rev{However, such an approach warrants a warning: as we have identified in our previous work~\cite{wintersgill2024law}, developers sometimes believe themselves to acquire a ``feel'' for licenses in this way despite misunderstanding their terms in some capacity, resulting in a false sense of confidence that can be a source of conflict with legal experts.}
\looseness=-1

Interviews also revealed how developers learned about unfamiliar licenses encountered during development. For example, \iresp{7} described ``[searching] mainly on the web, and looking at other cases. And also the basic Wikipedia$\dots$. [I once used] ChatGPT$\dots$ [for] interpreting the rules that are written [and] making sure that I understood well.'' Developers may also look for key terms in software licenses that pertain to them, as explained by \iresp{5}: ``Mostly what I'm looking for is, what are the restrictions on how to use it, and then, what are the particular attribution requirements or restrictions for it.'' The terms of interest can vary depending on the use case or organizational policies, such as in the case of \iresp{4}, who noted that ``as far as [my organization] goes, there is a strict policy for patent licensing and getting licenses for IP strategy.'' %

\finding{Developers report acquiring an understanding of licenses they encounter primarily through experience, despite the risk of misunderstanding them. To comprehend unfamiliar licenses, they often turn to online resources, including Generative AI tools and web search.}

\subsection{\ref{rq:3}: Licensing Tools}

To understand if and how developers use tools to assist with software licensing tasks, we asked participants to describe their use of licensing tools. \rev{We refer to a ``licensing tool'' as any software that can facilitate licensing tasks through automated analysis or by providing additional (licensing) context or information.}

\subsubsection{Tool adoption}
\label{sec:tool_adoption}

Of \completeresponses complete responses, only six participants indicated that they had used software tools to assist with software license compliance. This low adoption rate might suggest a lack of awareness or interest in licensing tools, despite the perceived difficulty of some licensing tasks, as discussed in \Cref{sec:difficult}. 

The interview participants elaborated, suggesting that the frequency with which licensing tasks are performed may be a factor in low tool use. \iresp{5} stated, ``I'm not aware of tools [because of] the fact that [licensing is] not something that I have to deal with every day...'' Similarly, \iresp{6} was unfamiliar with tools but mused that someone who engaged with licensing more might be more familiar: ``I am less aware [of licensing tools] ... [but] I would be completely unsurprised if there are established open source tools to assist in this [that] I am not personally aware of. But the people that work for me and do this when it is an important task very well may [know of them].''

\finding{Adoption of software licensing tools was low among respondents, who suggested that this may be due to the relatively low frequency at which licensing tasks are performed.}

\subsubsection{Tool benefits and shortcomings}
\label{sec:tool_benefits_shortcomings}

We asked those participants who indicated that they used licensing tools about their experiences with such tools to better understand how they are used in practice. 
First, respondents were asked to identify the benefits these tools provided. Given the small size of this group, no clear trends emerged, although we can extract some notable reasons for using such tools. Participants expressed that tools were useful in identifying the set of licenses present within software and its dependencies, including \resp{4}, who noted that tools were useful for ``automat[ing] recursive gathering of licenses,'' and \resp{43}, who appreciated tools that could increase ``[v]isibility on transitive dependencies'' and ``point out troubling dependencies.'' \resp{1} went a step further, claiming that tools help them ``ensure licenses are always compatible on every release.'' %

Tools can also help respondents choose licenses to assign to their software. In their survey response, \iresp{7} indicated that they benefited from tools that could give ``an overview of all licensing options [and support] the selection of the right option.'' They also noted a capability of ``[g]enerating terms that I can copy/paste,'' though it is not clear whether this referred to providing the text of existing licenses or creating new licenses. 
In interviews, \iresp{5} and \iresp{7} both referred to using choosealicense.com~\cite{choosealicense} to assist them when assigning licenses to software, praising its ability to ``answer your questions and then recommend the best license,'' as \iresp{7} put it.

We also asked the respondents to identify the difficulties or shortcomings that they had encountered with the tools. 
Two respondents indicated that they had encountered issues with licensing tools being cumbersome or difficult to use, with \resp{43} describing usability as ``poor.'' 
Others noted that tools were not able to account for user feedback: in their survey response, \iresp{7} called current tooling ``[n]ot interactive enough,'' and noted that it could be improved with an ``[i]nteractive decision tree'' or a ``[r]eliable chat bot.'' 
Other concerns included the clarity of tools' outputs, such as \resp{43}'s observation that the ``[a]ctual effect of a license is obscure[d] to developers,'' and \resp{37}'s opinion that tools ``[m]ay abstract away finer details.''

\finding{Developers appreciated tools' help in identifying licenses, highlighting dependencies which may cause licensing issues, and choosing appropriate licenses for their software. However, tools are hindered by low usability, low interactivity, and shallow, unclear analyses.}

\subsubsection{Areas requiring better tool support}
\label{sec:need_support}

Given the low reported tool adoption, we investigated in interviews which licensing tasks respondents believed could most benefit from better tool support. 

Interviewees noted that ways to better explain licenses and their terms would be beneficial. To illustrate this, \iresp{5} cited how ``GitHub [has a] drop-down list when you choose a license, but it doesn't tell you what [the licenses] mean.'' 
\iresp{3} echoed this sentiment, elaborating that interpreting the terms of software licenses was a challenge: ``[T]here is a gap between what the license gives you as $\dots$ content and [how] you can interpret it$\dots$ [You] can understand the text, but if you need to learn how to interpret that, that would be nice, a tool that gives you the interpretation.'' 

Similarly, \iresp{7} sought a tool to highlight the differences between licenses: ``$\dots$ I would like to have something to compare licenses. Because I think that there are some overlaps between some licenses, and sometimes it's hard to see the borders of the license.''

Beyond this, some interviewees expressed a need for tooling support for some of the more difficult compliance tasks, as defined in \Cref{fig:task_difficulty}. %
For example, \iresp{7} noted that they did not know of any tools to verify license compliance. %

Noting developers' difficulties in understanding software licenses, \iresp{4} identified a need for assistance in communicating with legal experts: ``[T]here [could] be a feature that can guide technical people like us, about the legal terminologies better [$\dots$] as well as the communication between the legal people and the technical people. [$\dots$] For example, what kind of things do we need to talk about?"
\looseness=-1

\finding{Developers expressed a desire for increased tool support to explain licenses and their terms, compare licenses, verify license compliance, and provide guidance on information to share with legal experts.}

Interviewees also discussed the potential impact of large language models (LLMs) on software licensing, though their views varied. Some, like \iresp{2}, envisioned useful applications: ``I think it would be pretty useful to have something where [you can ask] AI [$\dots$]: `I want a license with those terms', and it will say, `You can't use this exactly like this. So here's the licensing$\dots$' [or] give it an [upstream] repository, and it will read through [it] and then tell you, `[You] can use that,' or `[You] need to watch out on this part$\dots$'\hspace{0pt}'' This respondent described using AI to assist with modifying an existing license to create a custom license for their project: ``I used AI for a general overview because I very, very roughly knew what I wanted, and I [did not] want to spend forever searching for licens[es]. So I was like, `let's search which licenses are up there$\dots$' then [I can] just [manually] modify it slightly as I think is needed to do everything at a basic level that I want$\dots$'' Others, such as \iresp{5}, took a more cautious stance, noting that before using AI-based approaches, ``[it would be worth] looking into AI and seeing how good it is at describing those type of licensing situations$\dots$'' Still others, including \iresp{6}, viewed such approaches negatively: ``[M]y personal opinion is that [using generative AI for license creation] would be incredibly fraught, because [if you] let ChatGPT or Gemini or [another LLM] generate the legal agreements... then something will go wrong and you will get sued$\dots$''

\finding{Developers expressed mixed opinions on the potential of generative AI with respect to software licensing: the impacts it might have are not yet clear, and further investigation is required.} 

Taking a broader view, \iresp{5} suggested that tools would benefit from being built into platforms that developers already use regularly: ``[I]t could be something that is integrated into [an IDE such as] IntelliJ or GitHub that [prompts users] automatically$\dots$ before [they] publish this piece of code, [for example, by asking] what license would you want to choose?''

However, interviewees stressed that any tooling would need to be reliable, given the stakes. \iresp{3} noted that while ``I know there are tools for this$\dots$ [I want] something that you can trust 100 percent. I don't want to [find] false positives$\dots$'' Similarly, \iresp{7} cited a concern about making licensing mistakes: ``[I’m] worried about the licenses, so if we can have a tool that makes [licensing tasks] easier, less stressful$\dots$ because we are talking about the laws and sometimes I'm worried about that, I don't want to [make] mistakes$\dots$''

In general, licensing tools with sufficient capabilities can improve the process of performing licensing tasks in various ways, as long as they are sufficiently reliable and accessible.

\finding{Respondents emphasized that licensing tools should be reliable and readily accessible, particularly through seamless integration into existing development tools and platforms (\eg IDEs and GitHub).}

\section{Discussion and Implications}
\label{sec:discussion}
This section outlines the implications of our findings for practitioners and suggests areas that need additional focus. %

\subsection{Who is responsible for license compliance?}
\label{sec:who_responsible}

In \Cref{sec:roles}, we observed that respondents were divided about who exactly should be responsible for handling license compliance tasks, but 
the most common answer was that legal experts should ultimately be responsible. Consequently, in \Cref{sec:process}, we observed that when faced with licensing issues, 
respondents' most-reported strategy was indeed to escalate to legal experts. However, we also note that such experts are not always available to developers. %

This result is in contrast to the findings of our recent study regarding considered how legal practitioners approach software license compliance tasks \cite{wintersgill2024law}, which found that legal professionals believed that compliance should be a team effort, but that the primary responsibility for compliance should fall on developers, who are most familiar with the project. Legal professionals may have limited technical training and are not closely involved in the code base. In their view, it was the role of legal teams to provide developers with the tools, resources, and training they need to succeed and to provide guidance on only the more challenging issues.

Both developers and legal practitioners appear to believe that the ultimate responsibility for compliance should fall on the other party, and both provide valid arguments for why this should be the case. %
In reality, the answer likely lies somewhere in the middle, with an appropriate collaboration of legal and development teams that can bring the optimal combination of technical knowledge and legal background.  However, as noted in our previous work \cite{wintersgill2024law} and in \Cref{sec:need_support}, 
this balance can be difficult to strike, since legal experts and software developers may have different interests and goals. More research is needed to explore the interaction between roles and determine ways to facilitate effective collaboration. 
\rev{This can involve understanding what materials legal experts need from developers to effectively analyze potential issues, while simultaneously providing developers with tools to package that information in a way that is accessible and understandable to legal practitioners.}
\looseness=-1
\vspace{-10pt}
\looseness=-1

\subsection{Where to focus future efforts?}

Although our results show that some licensing tasks may be infrequent for developers, their importance is undiminished: failure to comply with licenses can result in significant costs~\cite{vizio_1, ai_copyright, johndeere, gunningham2004social, cost1, cost2}, justifying focused investment in licensing tools.%

As shown in \Cref{sec:common} and \Cref{sec:difficult}, developers perform software licensing tasks at different rates, and some tasks are more difficult than others. 
While verifying license compliance was reported as one of the development tasks developers perform most frequently, it was also cited as the most difficult task by far. Despite this, respondents were broadly unaware of tools capable of assisting with this task. \Cref{sec:tool_adoption} showed low tool adoption, mainly due to the lack of tools better integrated into the software development process. As suggested in \Cref{sec:need_support}, licensing tools built into an IDE or a commonly-used service like GitHub would make them easier to access for developers while also mitigating usability concerns highlighted in \Cref{sec:tool_benefits_shortcomings}. %
Increased accessibility may also mitigate the last-minute timing of resolving licensing issues, as identified in \Cref{sec:process}. 
\rev{The broad array of licensing factors considered (\Cref{sec:licensing_factors}), roles involved (\Cref{sec:roles}), and potential solutions to issues (\Cref{sec:process}) indicate that licensing tools must be able to take into account the context provided by the user and adapt to the circumstances of a given project.} %

\Cref{sec:need_support} also articulated developers' varying views on generative AI as it relates to software licensing. Our interviews revealed that while some are hopeful that AI-based tools can assist with software licensing in new and powerful ways, others expressed caution regarding their use for such a critical purpose. Therefore, we propose that an investigation into the suitability of generative AI for licensing tasks is warranted. An AI-based licensing tool would need to be capable of providing meaningful analyses of software licenses and software components' licensing status while also being reliable enough to avoid making inappropriate recommendations or providing services best left to legal practitioners.%

\section{Threats to Validity}
\label{sec:threats}
\textbf{Construct Validity.}
As with survey studies in general, our study cn only report participants' perceptions rather than actual practices. The interviews partially mitigated this threat, as they allowed participants to clarify and expand on the survey responses. However, studies such as ethnographic studies would be required to better understand developers' practices in the field.
\looseness=-1

\textbf{Internal Validity.} 
An additional possible threat is subjectivity in coding interviews' open-ended answers and transcripts. To mitigate such bias,  we utilized an open-coding methodology with a rigorous procedure to discuss and resolve conflicts. We employed diverse strategies to locate participants (professional networks via email and LinkedIn, Discord servers, snowball sampling) in an attempt to expand the pool to encompass different perspectives and minimize potential bias.
We followed best practices in formulating survey and interview questions, making sure that the questions were written clearly and concisely to avoid confusion and biased language, and conducted a small pilot study.  %
To encourage and facilitate participation from developers online, the survey and interviews were designed to be completed quickly:
the survey could be finished in about ten minutes, and interviews were limited to at most 30 minutes, meaning that not all of a participant's views were likely heard. We limited confirmation bias in qualitative analysis by independent coding, discussing disagreements, and arriving at a consensus based on the data.
\looseness=-1

\textbf{External Validity.} 
Considering our study's response rate and the potential for self-selection bias, the conclusions drawn in this study apply only to the population that participated in our survey and follow-up interviews.  Our results cannot be generalized to the larger population, but in light of the themes that emerged, we believe that other software developers likely share some of the same experiences and challenges. That said, our goal is not to claim generalizability but to illustrate the current state of practice and to identify potential pain points that hinder developers from achieving efficient and effective license compliance for their projects.

\section{Conclusion}

We present a study of current software licensing challenges faced by developers through a survey of \completeresponses developers and \interviewcount follow-up interviews, deriving \total{totalfindings} key findings encompassing: 
(i) how frequently different licensing tasks are performed; 
(ii) their difficulty; 
(iii) the roles developers perceive as addressing licensing issues, 
(iv) processes for resolving those issues; and
(v) the benefits and shortcomings of licensing tools.
\rev{These results reveal actionable insights for the creators of licensing tools by highlighting areas that need improvement, such as low perceived interactivity and shallow analyses, as well as identifying both common and difficult licensing tasks as areas that can most benefit from future license tool development.}

The respondents had different strategies for handling licensing issues and differing views on who is responsible for addressing them. Although many reported performing licensing tasks infrequently and without licensing tools, they also highlighted difficult tasks, particularly verifying license compliance. As such, we recommend that future research and tool development focus on pain points and investigate the suitability of generative AI for licensing tasks.

\balance
\bibliographystyle{ACM-Reference-Format}
\bibliography{references}

\end{document}